\documentclass[12p]{iopart}
\usepackage{import}
\usepackage[dvipsnames, table]{xcolor}
\usepackage{graphicx}
\usepackage{subcaption}
\usepackage{url}
\usepackage{cite}
\usepackage{amsfonts}
\usepackage{subcaption}
\usepackage{hyperref}
\usepackage{comment}
\expandafter\let\csname equation*\endcsname\relax
\expandafter\let\csname endequation*\endcsname\relax
\usepackage{amsmath}
\usepackage{adjustbox}
\makeatletter
\providecommand{\doi}[1]{%
  \begingroup
    \let\bibinfo\@secondoftwo
    \urlstyle{rm}%
    \href{http://dx.doi.org/#1}{%
      doi:\discretionary{}{}{}%
      \nolinkurl{#1}%
    }%
  \endgroup
}
\makeatother
\usepackage[nameinlink,noabbrev,capitalize]{cleveref}

\renewcommand{\cref}{\Cref}

\let\oldexp\exp
\renewcommand{\exp}[1]{\oldexp{\left( {#1} \right)}}

\begin{document}

\title{Axial HoloTile: Extended Depth-of-Focus of Dynamic Holographic Light Projections}
\author{Andreas Erik Gejl Madsen, Jesper Gl\"uckstad}

\address{SDU Centre for Photonics Engineering, University of Southern Denmark, Campusvej 55, 5230 Odense-M}
\ead{gejl@mci.sdu.dk, jegl@sdu.dk}

\begin{abstract}
This publication extends the HoloTile framework to three dimensions, introducing the ability to generate arbitrary dynamic patterns composed of extended depth-of-field non-diffractive beamlets with theoretically $100\%$ diffraction efficiency. In particular, we demonstrate experimentally the generation of speckle-reduced reconstruction patterns, consisting of spatially multiplexed extended Bessel-like beamlets, implemented on a phase-only spatial light modulator (SLM). 

Due to the inherent separation of the tiled subhologram and the point spread function shaping hologram in HoloTile, we show that the reconstruction amplitude can be expressed as a simple convolution of the contributions from the two holograms. This results in a discretely sampled reconstruction, with each spatial frequency component exhibiting long DoF with characteristic Bessel beam properties. This separation facilitates spatial and temporal multiplexing of both contributions, and allows for real-time dynamic patterning with extended DoF. Additionally, a geometric analysis is included, allowing for the direct calculation of the propagation characteristics of the beamlets.
\end{abstract}
\noindent{\it Keywords\/}: HoloTile, Holography, Propagation, 3D-printing, Volumetric Additive Manufacturing


\maketitle
\ioptwocol

\section{Background}

Dynamic manipulation of light in two and three dimensions is crucial for modern scientific research and practical applications, enhancing control over light-matter interactions. This capability drives advancements in neuroscience \cite{papagiakoumou_scanless_2010,papagiakoumou_optical_2013}, microbiology, optical manipulation \cite{grier_revolution_2003,rodrigo_real-time_2004,rodrigo_actuation_2005}, materials processing \cite{olsen_multibeam_2009}, microfabrication \cite{kawata_finer_2001,galajda_complex_2001}, and more.

This study extends the HoloTile \cite{madsen_holotile_2022,madsen_digital_2024,gluckstad_holotile_2024,gluckstad_holographic_2022,gluckstad_holographic_2023} framework by introducing extended depth-of-focus (DoF) beamshaping for increased axial control, enabling real-time dynamic light shaping of long-propagating beamlets. This advancement supports applications in volumetric additive manufacturing (VAM) \cite{kelly_volumetric_2019,loterie_high-resolution_2020,regehly_xolography_2020,shusteff_one-step_2017,madridwolff_controlling_2022,alvarez-castano_holographic_2024-1}, laser material processing (LMP) \cite{salter_adaptive_2019,liu_dynamic_2018,kontenis_optical_2022}, laser-based lithography \cite{wang_progresses_2021}, and optical coherence tomography (OCT) \cite{lee_bessel_2008,curatolo_quantifying_2016}. In particular, non-diffractive beams with extended depth-of-focus (DoF) such as pin beams \cite{zhang_robust_2019,yu_self-healing_2023}, needle beams \cite{grunwald_needle_2020}, and Bessel beams \cite{khonina_bessel_2020}, etc. are of significant interest in numerous areas due to their attractive properties for e.g., self-healing beamlets and parallel optical particle guidance along the optical axis \cite{alonzo_helico-conical_2005,daria_optical_2011}. In this publication, one instance of such beams are shown to be incorporated into the HoloTile framework; the Bessel beam.

Considering carefully the creation of HoloTile holograms; the combination of a tiled object hologram $H_\textrm{sub}$ and a separate PSF hologram $H_\textrm{PSF}$, allows for several interesting conclusions to be drawn. The hologram plane, as it is displayed on the phase only spatial light modulator (SLM) can be expressed as
\begin{equation}
    H_\textrm{SLM}(\xi, \eta ; t) = H_\textrm{PSF} \cdot \sum_{i,j=0}^{N_t} \delta(\xi - i \ell_s, \eta - j \ell_s)  \otimes H_\textrm{sub}
\end{equation}
where $\ell_s$ is the physical size of the subholograms and $\otimes$ denotes the colvolution operation. The tiling is expressed by the convolution of a comb array and the subhologram. From this, the field in the focal plane can be expressed directly by the optical Fourier transform of a convex lens or by free-space Fraunhofer propagation:
\begin{align}
    A_f(x, y; t) &\propto A_\textrm{PSF} \otimes \sum_{m,n \atop = -\infty}^{\infty} \delta\left(x - \frac{m}{\ell_s}, y - \frac{n}{\ell_s}\right) \mathcal{F}\{ H_\textrm{sub} \} \nonumber \\
     &\propto  A_\textrm{PSF}(x, y, z; t) \otimes  A_\textrm{sub}\left(x - \frac{m}{\ell_s}, y - \frac{n}{\ell_s}; t\right) \label{eq:conv}
\end{align}
where $A_\textrm{PSF}$ is the Fourier transform of the PSF shaping hologram, and $A_\textrm{sub}$ is the Fourier transform of the subhologram which, due to the tiling operation, is only sampled and thus defined in specific locations defined by the spatial frequency comb array $\sum\limits_{m,n = -\infty}^{\infty} \delta\left(x - \frac{m}{\ell_s}, y - \frac{n}{\ell_s}\right)$. 

The aperture of the SLM, the input beamshape and, in turn, the inherent optical point spread function (PSF), are included in $H_\textrm{PSF}$ and $A_\textrm{PSF}$, respectively. 
Thus, $|A_\textrm{sub}|^2$ defines an intensity pattern of regularly spaced points, which may be modulated freely, akin to enabling or disabling single pixels on a digital display. 
The PSF reconstruction $A_\textrm{PSF}$, however, governs the shape of each of the points in $A_\textrm{sub}$. 
Both the shape and location\footnote{Within the unit cell of each reconstruction point.} of the PSF can be modulated independently from $A_\textrm{sub}$. 
This simple separation of terms contributing to the final reconstruction allows for greatly increased illumination and aberration control. 
Importantly for this publication, the shape of the PSF naturally also extends axially.
This effectively means that any beam shape, axially extended or otherwise, can be multiplexed spatially and temporally in any given diffractive pattern, hence the dependence on $z$ and $t$ in $A_\textrm{PSF}$ and $t$ in $A_\textrm{sub}$. 
In the following sections, we demonstrate the generation of extended DoF Bessel beams, and their incorporation into the HoloTile framework.

\section{HoloTile Bessel Beam Generation}
\label{sec:bessel-gen}
The ring beam shaper defined in the HoloTile framework is given by the phase profile
\begin{equation}
\label{eq:ring-phase}
	\phi_\textrm{ring}(r) = -\beta r
\end{equation}

\begin{figure}
\begin{subfigure}[b]{\columnwidth}
	\centering
	\includegraphics{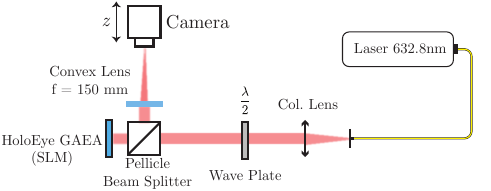}
	\caption{Lensed optical system. The SLM modulated field is optically Fourier transformed by a physical convex lens.}
	\label{fig:optical-setup}
\end{subfigure}
\begin{subfigure}[b]{\columnwidth}
	\centering
	\includegraphics{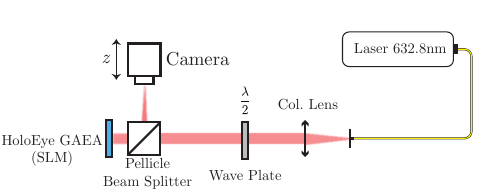}
	\caption{Lensless optical system. The SLM modulated field is optically Fourier transformed by a convex lens phase superimposed on the SLM.}
	\label{fig:optical-setup-lensless}
\end{subfigure}
\caption{The experimental setups for capturing lensed and lensless holographic reconstructions. A fiber coupled HeNe laser is collimated and incident on a phase-only SLM. The modulated reflected light is diverted in a beam splitter and captured directly in camera.}
\end{figure}

where $r$ is the radius from the optical axis in the hologram plane, $\beta = \frac{2 \pi r_0}{f \lambda}$, $r_0$ is the radius of the desired ring, $f$ is the focal length of the Fourier transforming lens, and $\lambda$ denotes wavelength. In the focal plane of the Fourier transform lens, the reconstruction shows, as expected, the shape defined by the subhologram $H_\textrm{sub}$ in which each point in the spatial frequency grid is shaped into well-defined rings.

\begin{figure*}
\centering
	\includegraphics{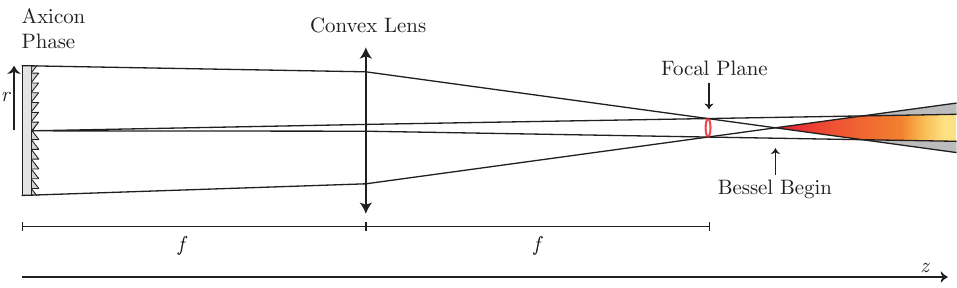}
	\caption{Illustration of the working principle of the ring-generated Bessel beams. A negative axicon phase diverts the incoming light, and is focused to a ring in the focal plane. Following a small propagation, the intersection of the rays from opposite sides of the axicon ring results in the generation of a Bessel beam (orange area). At a certain point, the side-lobes of the Bessel beam extend beyond the extent of the generating ring (gray area).}
	\label{fig:bessel-prop}
\end{figure*}

Interestingly, as the observation plane is moved axially along the propagation axis a specific distance from the Fourier plane, each ring converges to a point and continues to propagate as a Bessel-like beamlet \cite{schwarz_fabrication_2020}. As the propagation continues, the lobes of the Bessel beamlets expand to be larger than the original ring, at which point spatial intra-spectral interference can occur. A simplified illustration of the Bessel beam generation is shown in \cref{fig:bessel-prop}. Importantly, since the PSF shaping hologram directs all light into the sampled spatial frequency components, the reconstructions have a theoretical diffraction efficiency of $100\%$.

With the separation of pattern and output pixel shape expressed in \cref{eq:conv}, each point in the output pattern $A_\textrm{sub}$ is shaped, and will propagate, with Bessel beam characteristics.

\begin{figure*}
	\centering
	\includegraphics[trim={0.4cm, 1.4cm, 0cm, 2cm}, clip, width=\textwidth]{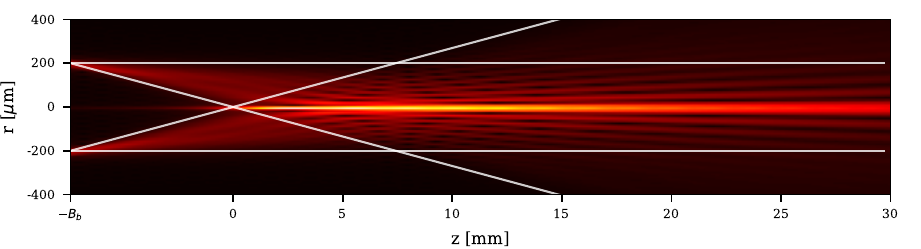}
	\caption{Field amplitude at increasing propagation distances from the beginning of the Bessel beam at $z=0$ mm to $z=30$ mm of single ring-generated Bessel beam. Overlayed is the rays resulting from the geometric analysis in \cref{sec:geometric}. The Bessel-generating ring is located at $z=-B_b$; the characteristic distance between the ring and the beginning of the Bessel beam.}
	\label{fig:bessel-prop-sim}
\end{figure*}

\subsection{Geometric Characterization}
\label{sec:geometric}
As previously remarked, there exists a characteristic distance between the focal plane in which the rings are clearly defined, and what might be considered the beginning of the Bessel beam propagation. Furthermore, there is also a limit to the propagation length over which the lateral extent of the Bessel beams can be considered smaller than the unit cell of HoloTile, i.e., the diameter of the Bessel generating rings. These characteristics can be approximated, following a 2D geometric analysis of the system in \cref{fig:bessel-prop}. Let the focal length of a convex lens be $f$, and place a negative axicon phase in accordance with \cref{eq:ring-phase} in the front focal plane of the lens. In the back focal plane, the characteristic ring pattern forms following the rotationally symmetric focus of the tilt of the axicon. Following the back focal plane, the ring defocuses, until the two rays originating from the outer radii of the axicon and lens intersect. From this point, Bessel beam interference emerges (red-yellow shaded area), as it would for an annular aperture. In the geometry, it is also apparent that when the innermost rays intersect the outer from the opposite side, light escapes the extent of the propagating Bessel beam (gray shaded area), thus giving rise to interference between adjacent propagating Bessel beams.

\begin{figure*}
	\centering
	\includegraphics[trim={0cm, 0cm, 0.1cm, 0.1cm}, clip, width=\textwidth]{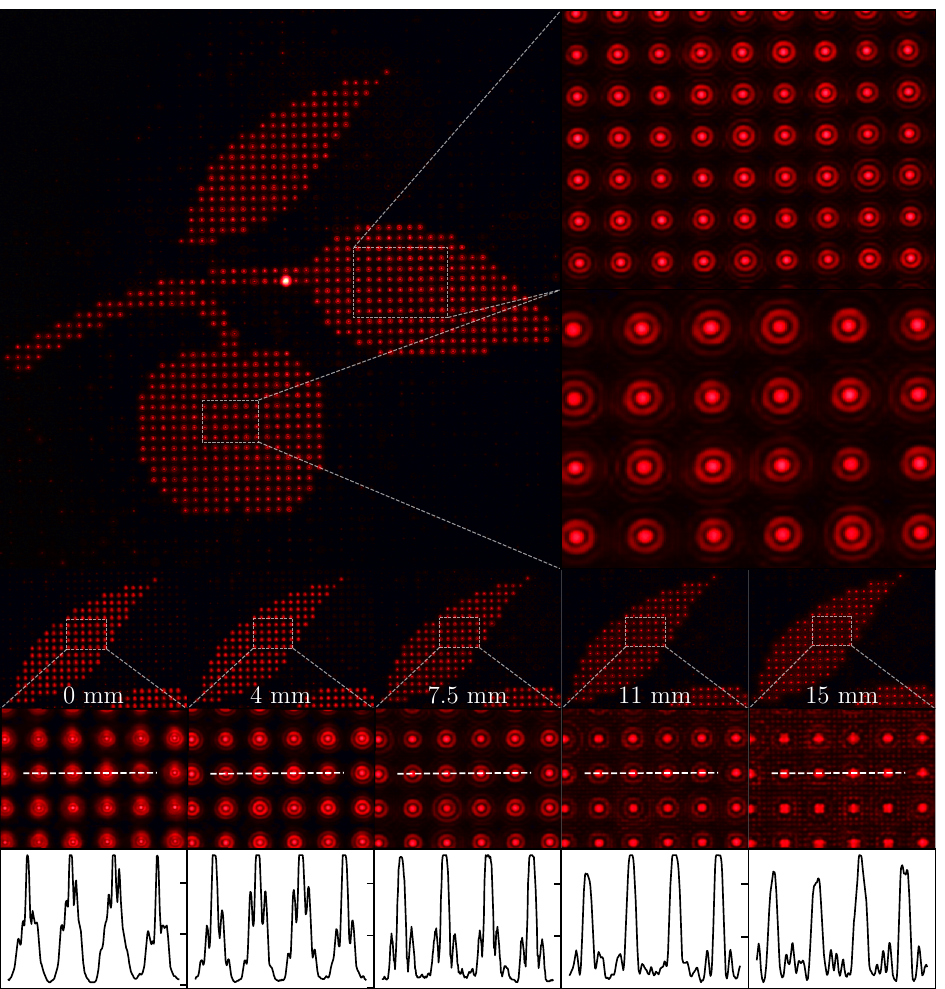}
	\caption{}
	\label{fig:lensed-comp}
\end{figure*}

\begin{figure*}
	\centering
	\includegraphics[trim={0cm, 0cm, 0.1cm, 0.1cm}, clip, width=\textwidth]{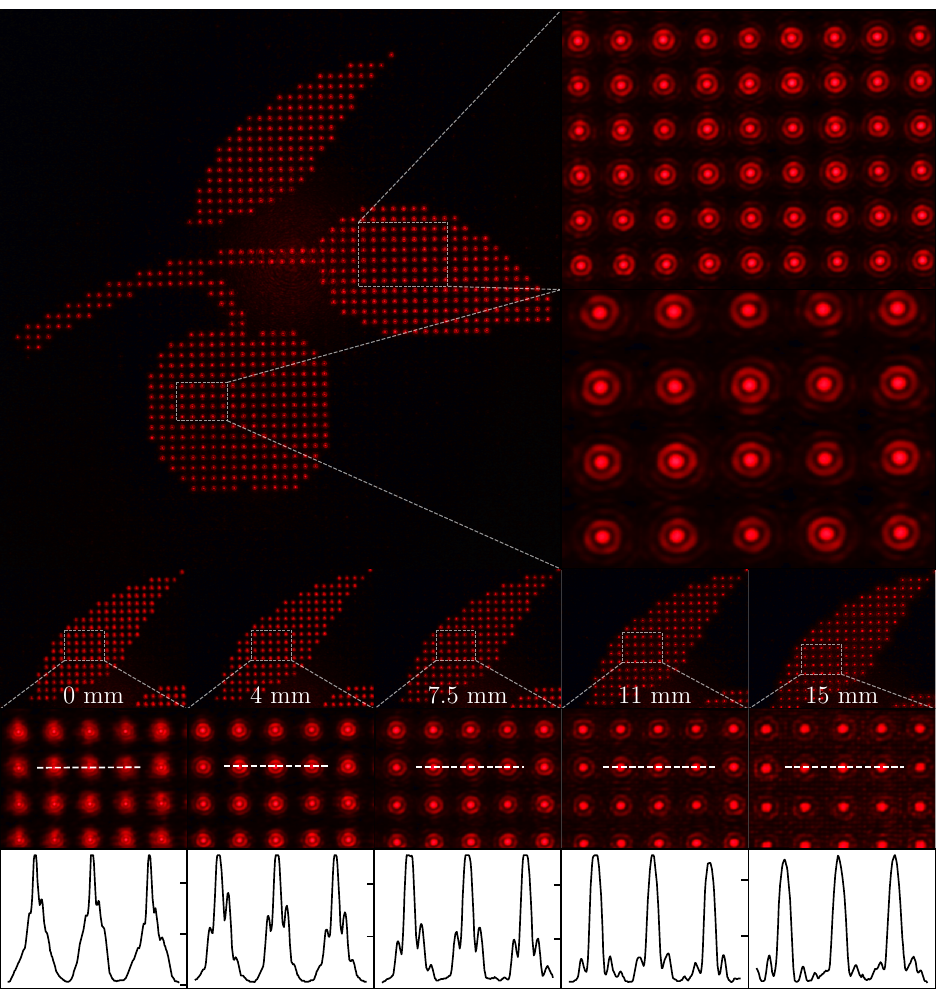}
	\caption{}
	\label{fig:lensless-comp}
\end{figure*}

The beginning of the Bessel beam is thus given as the z-coordinate at the intersection between the outermost rays. To express the Bessel beam beginning and its effective propagation length, we can approximate the system using the ABCD matrix approach. The final height and angle of the rays in the system is given by
\begin{equation}
\label{eq:abcd-full}
	\begin{bmatrix}
           r_f \\
           \theta_f \\
         \end{bmatrix}
      = 
      	\begin{bmatrix}
           1 & d \\
           0 & 1 \\
         \end{bmatrix}
         \begin{bmatrix}
           1 & f \\
           0 & 1 \\
         \end{bmatrix}
         \begin{bmatrix}
           1 & 0 \\
           -1/f & 1 \\
         \end{bmatrix}
         \begin{bmatrix}
           1 & f \\
           0 & 1 \\
         \end{bmatrix}
         \begin{bmatrix}
           r_\textrm{SLM} \\
           \theta_0 \\
         \end{bmatrix}
\end{equation}
where $r_\textrm{SLM}$ is the characteristic radius of the SLM, $d$ is the free-space propagation following the focal plane, and $\theta_0$ is the deflection angle of the light due to the axicon phase;
\begin{align}
	\theta_0 &= \arctan\left( \frac{\lambda \beta}{2\pi} \right) \nonumber \\
	&= \arctan \left( \frac{r_0}{f} \right)
\end{align}
The resulting rays of this geometric model are overlayed on a wave-based propagation simulation in \cref{fig:bessel-prop-sim}.
Simplifying \cref{eq:abcd-full} yields the final expressions for the outermost ray height and angle :
\begin{align}
	r_f &= f\theta_0 - \frac{d r_\textrm{SLM}}{f} \\
	\theta_f &= -\frac{r_\textrm{SLM}}{f}
\end{align}
Now, to find the beginning of the Bessel beam, $d$ is isolated in $r_f(r_\textrm{SLM}, \theta_0) = 0$, and can be expressed as
\begin{equation}
	B_b = d = \frac{f^2 \theta_0}{r_\textrm{SLM}}
\end{equation}
Similarly, the point at which the lateral extent of the Bessel beam exceeds the generating ring is found by expressing $r_f(r_\textrm{SLM}, \theta_0) = r_f(0, -\theta_0)$ and isolating $d$:
\begin{equation}
	B_e = d = \frac{2f^2 \theta_0}{r_\textrm{SLM}}
\end{equation}
Which results in a propagation length of:
\begin{align}
	B_\ell = B_e - B_b &=  \frac{f^2 \theta_0}{r_\textrm{SLM}} \nonumber  \\
	&= \frac{f^2}{r_\textrm{SLM}} \arctan \left( \frac{r_0}{f} \right)
\end{align}
Hence, there exists a non-linear relationship between the radius of the Bessel-generating rings $r_0$ and the propagation length $B_\ell$.
Although the side lobes of the Bessel beam extend beyond the ring diameter after this point, they are still dim compared to the central lobe. Therefore, the practical propagation length until significant spatial intra-spectral interference is observed is significantly larger, as is observed in \cref{fig:bessel-prop-sim,fig:lensed-comp,fig:lensless-comp}.
This result can be tied back to the original HoloTile equations \cite{madsen_holotile_2022} in order to express the required number of tiles on the SLM, and thus the required sub-hologram resolution $m_\textrm{sub}$, for a given propagation distance $B_\ell$, without the Bessel-generating rings interfering with neighbouring rings. For the rings to not extend beyond the unit cell of each spatial frequency component, the following inequality can be expressed \cite{madsen_holotile_2022}:
\begin{equation}
	r_0 \leq \frac{N_t \lambda f}{4 r_\textrm{SLM}}
\end{equation}
Where $N_t$ is the number of tiles on a square SLM. This requires that
\begin{align}
	N_t \geq \frac{4r_\textrm{SLM}}{\lambda} & \tan\left( \frac{B_\ell r_\textrm{SLM}}{f^2} \right) \\
	&\Downarrow \nonumber \\ 
	m_\textrm{sub} &\leq \frac{2r_\textrm{SLM}}{N_t \ell_p} 
\end{align}
Effectively tying the HoloTile subhologram tiling with the propagation of non-diffractive Bessel beamlets.

\section{Experimental Results}
To validate the propagation of the axicon-generated Bessel beams, a simple Fourier holography setup is employed, as illustrated in \cref{fig:optical-setup}. A fiber-coupled 632nm HeNe laser is collimated and modulated by a HoloEye GAEA phase-only Spatial Light Modulator\cite{noauthor_gaea-2_2022}, after which the resultant spatially Fourier transformed patterns are captured by a camera sensor (Canon M6 Mark II) that can be translated axially by an electronic stage. A subhologram corresponding to tile number $N_t = 20$ \cite{madsen_holotile_2022} of the University of Southern Denmark tree branch logo is calculated by an adaptive weighted Gerchberg-Saxton algorithm \cite{wu_adaptive_2021}, and is then tiled to form a $2160\times 2160$ hologram. Then, the separate $2160\times 2160$ PSF shaping hologram is calculated using \cref{eq:ring-phase}, with $r_\textrm{SLM}=117.5\mu$m \cite{madsen_holotile_2022}. While the hologram is displayed on the SLM, the camera is alternately capturing photos and moving along the optical axis.

In \cref{fig:lensed-comp}, a summary of the results of the experiment is shown. From the top, the large capture and its insets demonstrate the grid effect associated with HoloTile. Each individual spatial frequency component has been shaped by the PSF shaping hologram. As expected, each component shows the distinct radial oscillations of a Bessel beam. Below, the propagation of the generated Bessel beams is shown. Captures from five different planes spanning $15$ mm, along with insets and line profiles through the Bessel beams, demonstrate the extended propagation of the beams. Additionally, at propagation distances $z=11.5$ mm and $z=15$ mm, the beginning of the spatial intra-spectral interference can be observed. However, as was previously noted, the central lobe remains bright compared to the interference.

It is evident in these reconstructions that the Bessel beam PSF shaping allows for extended propagation of the desired pattern. Over the whole propagation distance, the individual output pixels are well defined, and the variation in brightness is minimal.

HoloTile functions in a lensless configuration as well, by superimposing the phase profile of an equivalent Fourier transforming lens on the SLM. As such, the optical setup in \cref{fig:optical-setup} is altered by removing the second lens, and moving the camera and stage to the appropriate focus of $f_\textrm{slm} = 150$mm. The capture procedure is then repeated, and the results are shown in \cref{fig:lensless-comp}. Again, the Bessel beam shaped reconstructions show good likeness to the pattern throughout the propagation distance, with minimal spatial intra-spectral interference.

It is important to note the enhanced performance achieved by integrating extended Bessel beams with the HoloTile framework, as opposed to simply superimposing a Bessel beam generating phase on a conventional computer-generated hologram (CGH). In a conventional CGH without tiling, the individual spatial frequency components are indistinguishable from one another. Since any beam with extended axial propagation inherently possesses lateral extent, the overlapping and interference of individual point spread functions (PSFs) create inhomogeneities both axially and laterally. By separating the spatial frequency components to a freely chosen degree, the HoloTile framework enables clear and uninterrupted propagation of these beams.

\section{Conclusion}
The HoloTile framework achieves extended axial control through the generation of extended DoF Bessel beams, independent of target hologram calculations. This independence enables the real-time axial extension of arbitrary patterns using a spatial light modulator.

The extended Bessel beams have been characterized both geometrically and through simulations. Experimental validations in both lensed and lensless configurations demonstrate extended propagation of arbitrary patterns, remaining in focus over a distance of $z=15$ mm. The inclusion of Bessel beams is simply one instantiation of extended DoF HoloTile, and future versions may show various other modalities, e.g., pin beams, needle beams, etc. This advancement opens new possibilities for precise and efficient light shaping in various scientific and industrial applications.

\ack
This work has been supported by the Novo Nordisk Foundation, Denmark (Grand Challenge Program; \\ NNF16OC0021948) and the Innovation Fund Denmark.

\section*{References}
\bibliographystyle{ieeetr}
\bibliography{references}

\end{document}